\documentclass[11pt,a4paper]{article}

\usepackage[margin=2cm]{geometry}

\usepackage{cite}
\usepackage{amsmath,amssymb,amsfonts}
\usepackage{algorithm}
\usepackage{algpseudocode} 
\usepackage{graphicx}
\usepackage{textcomp}
\usepackage{comment}
\usepackage{todonotes}
\usepackage{booktabs}
\usepackage[caption=false,font=footnotesize]{subfig}
\usepackage{xcolor}
\usepackage[hidelinks, bookmarks=false]{hyperref} 
\usepackage{mathrsfs}
\usepackage[inkscapeexe=/Applications/Inkscape.app/Contents/MacOS/inkscape,inkscapearea=page]{svg}
\usepackage{array}
\def\BibTeX{{\rm B\kern-.05em{\sc i\kern-.025em b}\kern-.08em
    T\kern-.1667em\lower.7ex\hbox{E}\kern-.125emX}}

\usepackage{letltxmacro}
\usepackage{xparse}
\usepackage{multirow}

\AtBeginDocument{%
	\LetLtxMacro\autoreforig\autoref
	\RenewDocumentCommand{\autoref}{som}{%
		\IfBooleanTF{#1}{%
			\autoreforig*{#3}%
		}{%
			\autoreforig{#3}%
		}%
		\IfValueT{#2}{#2}%
	}
}
\def\equationautorefname~#1\null{%
	Eq.~(#1)\null
}

\newcommand{\sample}{x}
\newcommand{\sampleAfterTransformation}{y}
\newcommand{\width}{W}
\newcommand{\errorAbs}{\Delta}
\newcommand{\errorPerc}{\delta}
\newcommand{\errorBoundAbs}{E^\Delta}
\newcommand{\errorBoundPerc}{E^\delta}
\newcommand{\dataset}{\mathscr{D}}
\newcommand{\A}{\mathcal{A}}
\newcommand{\M}{\mathcal{M}}

\definecolor{lgray}{gray}{0.7}
\definecolor{llgray}{gray}{0.83}
\definecolor{lllgray}{gray}{0.92}

\newcommand{\boxForTable}[1]{\setlength{\fboxsep}{0pt}\framebox{#1\strut}}
\newcommand{\noBoxForTable}[1]{\setlength{\fboxsep}{0pt}\mbox{#1\strut}}
\newcommand{\firstPattern}[1]{\textcolor{red}{\textbf{#1}}}
\definecolor{greedyColor}{HTML}{404E5C}
\definecolor{bzipColor}{HTML}{35CE8D}
\definecolor{lz4Color}{HTML}{B7C3F3}
\definecolor{zstdColor}{HTML}{DD7596}
\definecolor{lossless}{HTML}{CA895F}

\newcolumntype{C}{>{\centering}p{0cm}}

\begin{document}

\title{Change a Bit to save Bytes: Compression for Floating Point Time-Series Data\\
\thanks{This work was supported by the IoTalentum Project within the Framework of Marie Skłodowska-Curie Actions Innovative Training Networks (ITN)-European Training Networks (ETN), which is funded by the European Union Horizon 2020 Research and Innovation Program under Grant 953442.}
}

\author{
	Francesco Taurone, Daniel E. Lucani, Marcell Fehér and Qi Zhang\\
	DIGIT, Department of Electrical and Computer Engineering \\
	Aarhus University \\
	\{francesco.taurone, daniel.lucani, sw0rdf1sh, qz\}@ece.au.dk
}
\maketitle

\begin{abstract}
	The number of IoT devices is expected to continue its dramatic growth in the coming years and, with it, a growth in the amount of data to be transmitted, processed and stored. Compression techniques that support analytics directly on the compressed data could pave the way for systems to scale efficiently to these growing demands. This paper proposes two novel methods for preprocessing a stream of floating point data to improve the compression capabilities of various IoT data compressors. In particular, these techniques are shown to be helpful with recent compressors that allow for random access and analytics while maintaining good compression. Our techniques improve compression with reductions up to 80\% when allowing for at most 1\% of recovery error.
\end{abstract}


\section{Introduction}
IoT devices generate large amounts of data that need to be transmitted, stored and analyzed. The advantages of compressing chunks of data include reductions in data transmission costs and more efficient use of bandwidth, as well as reduced data storage. We focus on time-series data compression algorithms \cite{TimeSeriesSurvey}, where the most common objective is to generate, packetize, compress and store a continuous data stream, it being IoT, financial, or for other use cases, using little memory and computing power.

\begin{figure}[tb]
	\centering
	\includegraphics[width = 0.7\columnwidth]{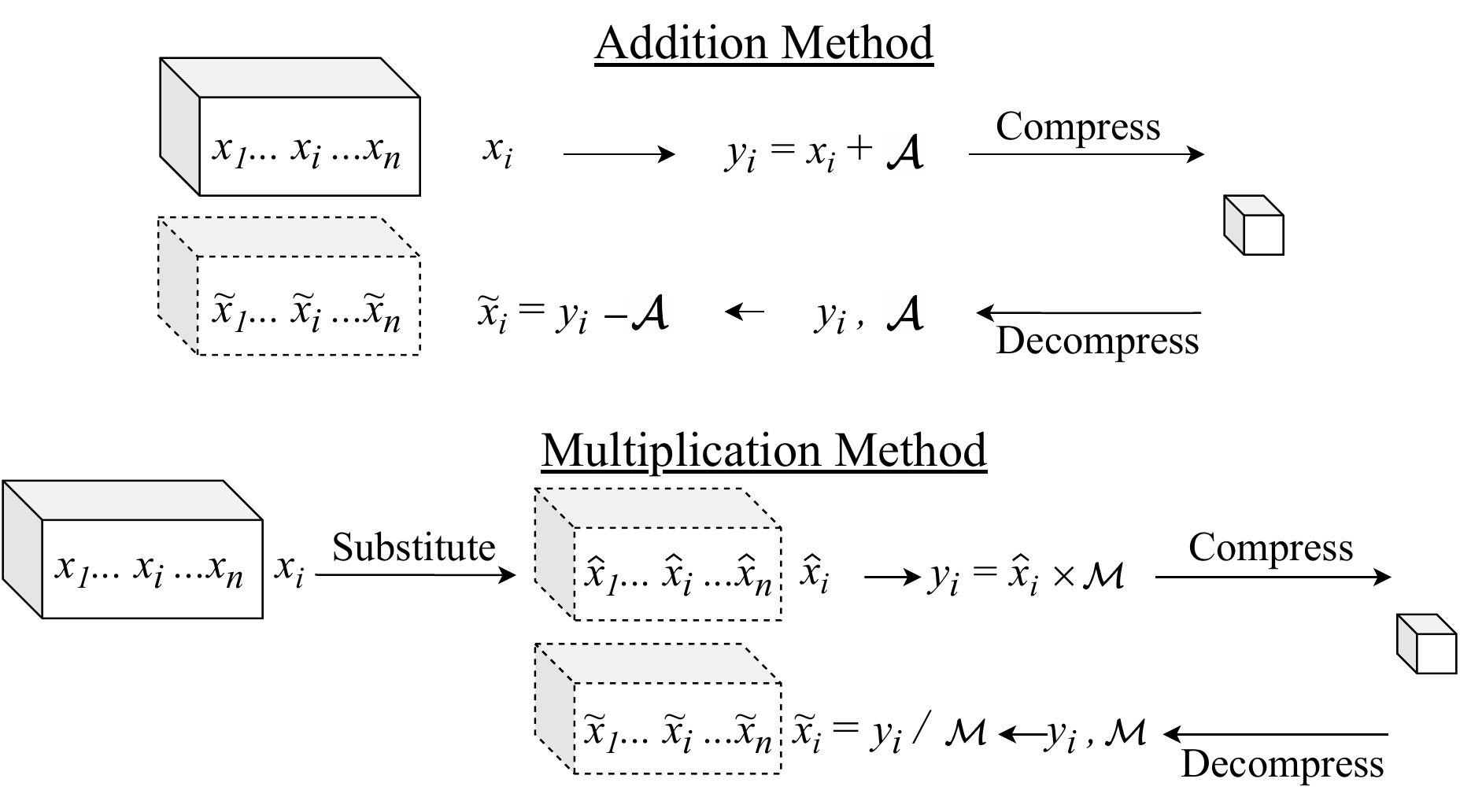} 
	\vspace{-0.7cm}
	\caption{Preprocessing and compression diagram from original $\sample_i$ to recovered $\tilde{\sample}_i$ with the two proposed methods. $\sample_i$ and $\tilde{\sample}_i$ might differ from each other.}
	\label{fig:compressionDiagram}
\end{figure}

\begin{figure}[tb]
	\vspace{-0.4cm}
	\centering
	\includegraphics[width = 0.7\columnwidth]{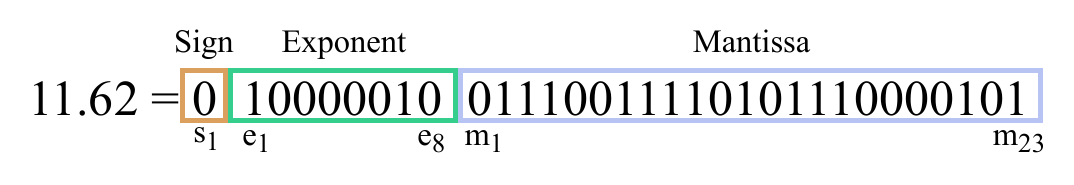}
	\vspace{-0.3cm}
	\caption{Encoding of a 32-bit IEEE 754 floating point number.}
	\vspace{-0.5cm}
	\label{fig:FPStructureExample}
\end{figure}

This paper focuses on data manipulations prior to using a compression algorithm. Our goal is to modify individual samples to make them better suited for compression than the original data stream. More specifically, we propose two novel transforms:  the \emph{addition} and \emph{multiplication transform}. We summarize them in \autoref{fig:compressionDiagram}, where assuming a one-dimensional set of samples [$\sample_1,\dots,\sample_n$] as input, after applying the transforms, we obtain a dataset [$\sampleAfterTransformation_1,\dots,\sampleAfterTransformation_n$].  We operate on a per-value basis exploiting floating point representation characteristics, whose basic structure is in \autoref{fig:FPStructureExample}, in order to cater to random access compressors, e.g., generalized deduplication\cite{GD}.  The parameters of each transformation are selected so that all $\sampleAfterTransformation_i$ have some identical bits at the same position (e.g., all zero bits in the last 21 mantissa bits in the last row of \autoref{fig:motivation}). In data recovery, the inverse transformation results in [$\tilde{\sample_1},\dots,\tilde{\sample}_n$], which is the reconstructed dataset. Although our methods are lossy, the user can specify an upper bound for the maximum recovery error (i.e., distortion) of each sample in the data stream. The larger the maximum allowed error, the more potential for compression.
\subsection{Motivating example}
Consider that we have two numbers $\sample_1$ and $\sample_2$ as in \autoref{fig:motivation}. Although they share a lot of digits in their decimal form, their mantissas have only 2 bits in common. When we first apply the addition transform, we select a parameter $\A$ to be added to all data in the stream, represented by $\sample_1$ and $\sample_2$ in this example. Here, we use $\A = 1738.0$, that generates $\sampleAfterTransformation_1 = 1791.333$ and $\sampleAfterTransformation_2 = 2047.333$. Although the transform seems simple, it results in multiple bits having the same value in both $\sampleAfterTransformation_1$ and $\sampleAfterTransformation_2$, which can then be compressed more effectively than the originals. Moreover, the exponents of both numbers are now the same as well.

For the multiplication transform, we first modify the bit streams judiciously to create patterns in the mantissa that, when multiplied by a specific $\M$ value, generate a sequence of zeros in the resulting mantissas. The first bit pattern transformation outputs $\hat{\sample}_1$ from $\sample_1$ and $\hat{\sample}_2$ from $\sample_2$ . We then multiply by $\M = 3.0$ to generate $\sampleAfterTransformation_1$ and $\sampleAfterTransformation_2$. We note that all 23 bits in the mantissa are common and could lead to more effective compression.
%

\begin{table}[t]
	\caption{
		Before and after preprocessing}
	\centering
	\begin{tabular}{ll}
		\toprule
		\multicolumn{1}{l}{\textbf{Original}} & Floating point representation (Sign$\cdot$Exp$\cdot$Mant)\\
		\hline
		\multicolumn{1}{l}{$\sample_1$ = 53.333} &
		1$\cdot2^5\cdot$[1.\boxForTable{1010101010101010101}0100]$_{\text{base2}}$ \rule{0pt}{2.5ex} \\
		\multicolumn{1}{l}{$\sample_2$ = 309.333} &
		1$\cdot2^8\cdot$[1.001\boxForTable{1010101010101010101}1]$_{\text{base2}}$ \\
		\textit{Common bits:}&\;\;\;\;\;\;\;\;\;\;\;\; xx\\
		\toprule
		\multicolumn{2}{l}{\textbf{Addition transformation, \boldmath$\A = 1738.0$} }\\
		\hline
		\multicolumn{1}{l}{$\sample_1 + \A$  = 1791.333} &
		1$\cdot2^{10}\cdot$[1.10\boxForTable{11111111}\boxForTable{01010101010}00]$_{\text{base2}}$ \rule{0pt}{2.5ex}\\
		\multicolumn{1}{l}{$\sample_2+\A$ = 2047.333} &
		1$\cdot2^{10}\cdot$[1.11\boxForTable{11111111}\boxForTable{01010101010}00]$_{\text{base2}}$ \\
		\textit{Common bits:}&\;\;\;x\;\;\;\;\;\;\;\;\,x\;xxxxxxxxxxxxxxxxxxxx\\
		\toprule
		\multicolumn{2}{l}{\textbf{Multiplication transformation, \boldmath$\M = 3.0$ }}\\ 
		\hline
		\multicolumn{2}{l}{$\hat{\sample_1}$ = 53.33333206176758, $\hat{\sample_2}$ = 309.3333435058594 } \rule{0pt}{2.5ex}\\
		\multicolumn{1}{l}{$\hat{\sample_1} \cdot \M$ = 160.0} &
		1$\cdot2^7\cdot$[1.01\boxForTable{000000000000000000000}]$_{\text{base2}}$ \\ 
		\multicolumn{1}{l}{$\hat{\sample_2} \cdot \M$ = 928.0} &
		1$\cdot2^9\cdot$[1.01\boxForTable{000000000000000000000}]$_{\text{base2}}$\\
		\textit{Common bits:}&\;\;\;\;\;\;\;\;\;\;\;\,xxxxxxxxxxxxxxxxxxxxxxx\\
		\bottomrule
	\end{tabular}
\vspace{-0.2cm}
\label{fig:motivation}
\end{table}
\subsection{Related work}
Exploiting data structure and its contextual meaning to achieve compression reductions has been researched extensively, e.g. with images, audio and time-series data \cite{TimeSeriesSurvey}. One of the crucial steps for compression is the initial dataset manipulation, where the objective is to clean it and filter out noise, outliers or other unwanted components that might undermine compression. This process is particularly crucial for data coming from IoT devices since sensors can be quite noisy and degrade over time \cite{sensorProblems}. It is often up to the user's knowledge and experience to prepare data for compression. A complementary approach, usually part of the compressor pipelines, is to represent the same information using predictors or models of the sampled system \cite{ISO15948}. Alternatively, we can change the domain of the signal to encode other characteristics, e.g frequency domain in DCT and DWT \cite{dct}. Some of these techniques perform well when applied to specific types of dataset, as part of specialized compressors pipelines. We propose two more general-purpose preprocessing methods that transform floating point time-series streams before passing them as input to a wide range of possible compressors. They improve the compressor effectiveness by increasing the number of bit values shared in the dataset to reduce its entropy. This objective is also part of the preprocessing method proposed by Klöwer et al. in \textit{Nature Computational Science}\cite{infoPreprocessing}, which we use as benchmark for the evaluation of our proposed techniques.

\section{Background}
\subsection{Performance metrics}
We ultimately want to show that these techniques improve compression performances in terms of output package size. We use the \textit{compression ratio} (CR) as metric, defined as
\begin{equation}
	\label{eq:compressionRatio}
	\text{CR} = \frac{\text{size of compressed data}}{\text{size of uncompressed data}}.
\end{equation}
We also need a parameter called \emph{maximum recovery error}, so the user can impose the recovery error upper bound, which ultimately depends on its application needs. We do so by defining it both in terms of \emph{absolute error}
\begin{equation}
		\errorBoundAbs =\max_{i} \errorAbs_{i} \quad \textrm{with  } \errorAbs_{i} =\lvert \tilde{\sample}_i - \sample_i  \rvert, \\
		\label{eq:errorUpperBoundAbs}
\end{equation}
 and \emph{relative} to the original value
\begin{equation}
	\errorBoundPerc = \max_{i} \errorPerc_{i} \quad \textrm{with  } \errorPerc_{i} =(\lvert \frac{\tilde{\sample}_i - \sample_i }{\sample_i}\rvert).
	\label{eq:errorUpperBoundRel}
\end{equation}

\subsection{Floating point numbers \text{(FP)}}

While there are various ways to represent real and rational numbers \cite{FPHandbook}, we focus on the standard IEEE-754 \cite{IEEE754}. Although our description and experiments use the 32-bit version, the proposed transforms are easily adapted to smaller or extended precision formats in the standard, e.g. 16, 64, and 128 bits. Given a FP number $n$, its 32-bit structure is divided into three parts, as per \autoref{fig:FPStructureExample}:
\begin{itemize}
	\item \textit{Sign `S'}: $\{s_1\}$ bit. `$0$' for $n\geq0$, `$1$' for $n<0$.
	\item \textit{Exponent `E'}: $\{e_1 \dots e_8\}$ bits. It is biased, meaning that it is interpreted as an unsigned integer once the bias $b=2^{k-1}-1 = 127$ is subtracted from it, where $k$ is the number of exponent bits.
	\item \textit{Mantissa `M'}: $\{m_1 \dots m_i \dots m_{23}\}$ bits. Each $m_i$ represents $2^{E -b - i}$.
\end{itemize} 

\begin{figure}[!t]
\centering
\includegraphics[width =0.7 \columnwidth]{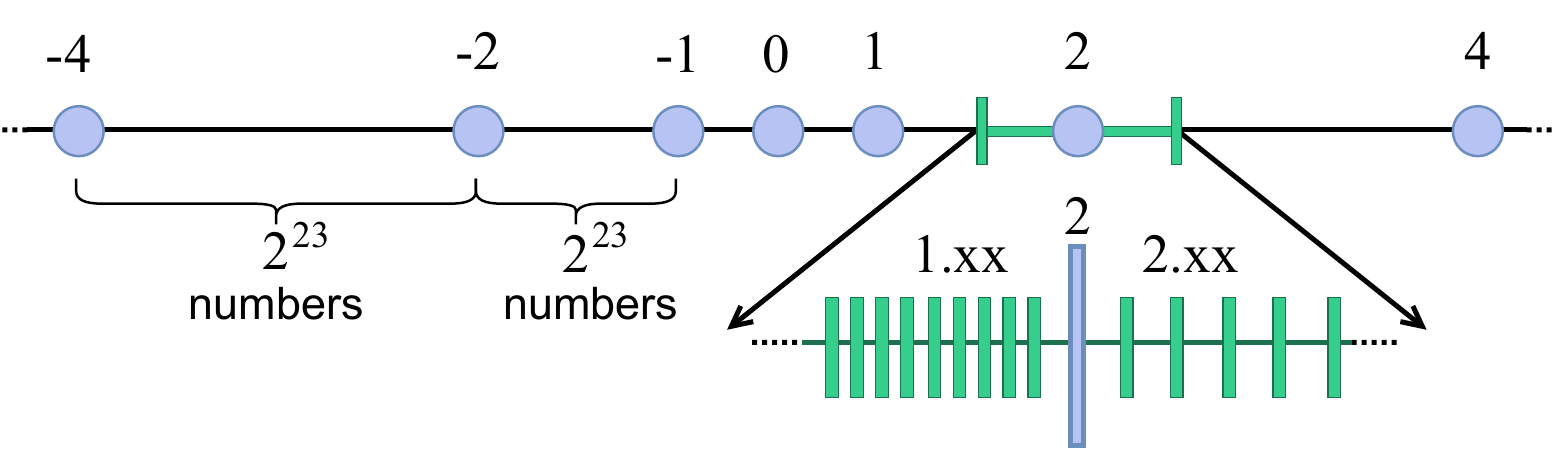}
\vspace{-0.7cm}
\caption{
	Floating point distribution on the real axis for 32-bits floats. 
}
\vspace{-0.3cm}
\label{fig:sequentialFP}
\end{figure}

In order to translate bits to numbers, we use the equations
\begin{align}
	\begin{split}
		n &= (-1)^S \cdot 2^{E- b} \cdot (1 + \textit{M} \cdot 2^{-23})\\
		&= (-1)^S \cdot [2^{E_U} +  m_1 \cdot 2^{E_U - 1}  +  \dots +  m_{23} \cdot 2^{E_U - 23}  ]
	\end{split}
	\label{FPFormula}
\end{align}
where $E_U = E - b \neq 0$. For each $n$, the smallest quantity we can use in its representation is {$2^{E_U - 23} $}, which depends on the exponent of the number. We can define this \textit{Precision} as
\begin{equation}
	P(E_U) = 2^{E_U-23}.
	\label{eq:precision}
\end{equation}
Therefore, two consecutive FP-representable real numbers in the region with precision $P(E_U)$, differ by $P(E_U)$, or more formally
\begin{align}
	\begin{split}
&\lvert n_1 - n_2\rvert \geq P(E_U),\forall n_1 \neq n_2 \in [2^{E_U}; 2^{E_U + 1}].\\
	\end{split}
\label{eq:consecutiveFPnumbers}
\end{align}
Since the distance between two consecutive FP-representable numbers depends on their exponent, i.e. $E_U$, the further we go from zero, the sparser floating point numbers become on the real axis, as per \autoref{fig:sequentialFP}.
When $n \in [0, 2^{-b}[$, with $b = 127$ for 32-bit floats, $E=0$ and the number is called \emph{subnormal}. The standard defines specific rules for representing and operating with these extremely small and uncommon values, and we will not treat them in the following. The only exception is $n = 0.0$, which is handled separately by the proposed transformations.

%
%
 
\section{Preprocessing methods}
\subsection{Addition transform}
\begin{figure}[!t]
	\centering
	\vspace{0.1cm}
	\includegraphics[width = \columnwidth]{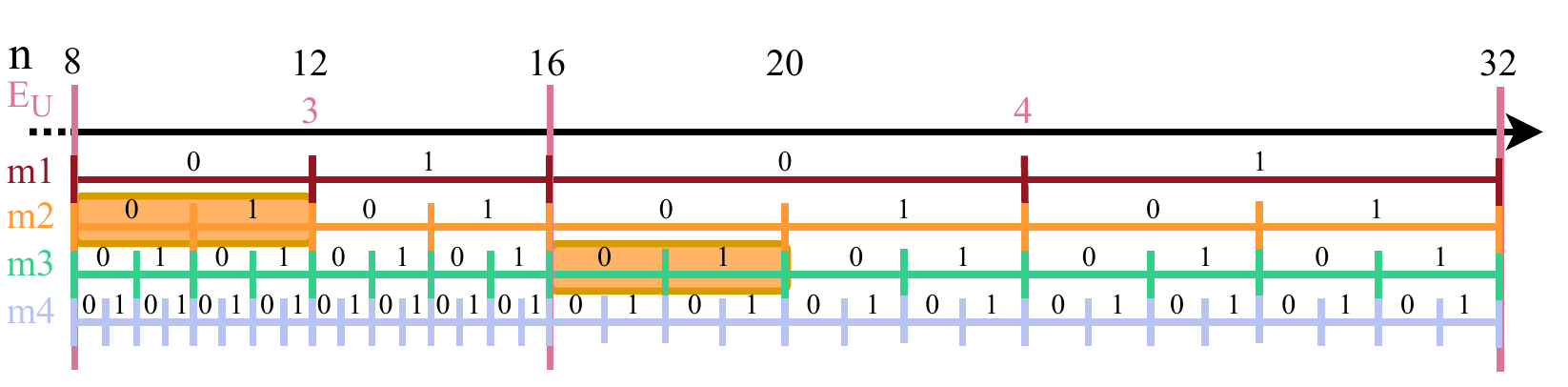}
	\caption{
		Dataset (range \textit{in orange}) with $\sample_i\in [8,12] $. By applying $\sampleAfterTransformation_i  = \sample_i+ 8$, we move the dataset to  the region $[16, 20]$, where numbers have larger exponent ($E_U = 4$, against the original $E_U = 3$). Therefore, every number in the shifted dataset will have $m_2 = 0$.
	}
	\label{fig:additionGuaranteedCommonBits}
	\vspace{-0.5cm}
\end{figure}

This method can be summarized as the following operation:
\begin{equation}
	\sampleAfterTransformation_i = \sample_i + \A , \quad\forall \sample_i \in \dataset \\
	\label{eq:formulaAdditionMethod}
\end{equation}
where $\A > 0$ is called \emph{addition parameter}, $\dataset$ is the dataset. The idea is to shift all samples of the dataset to a suitable region on the real axis. As illustrated in \autoref{fig:additionGuaranteedCommonBits}, by strategically choosing $\A$ we can guarantee that all numbers in dataset after the addition transform share the exponent and several mantissa bits. The cost is a possible recovery error due to changing precision from the original region of each sample $\sample_i$ to the new $\sampleAfterTransformation_i$’s precision.
In order to recover the original data, we apply the inverse transformation $\tilde{\sample}_i = \sampleAfterTransformation_i - \A$, thus $\A$ needs to be stored as metadata.

\subsubsection{Recovery error}
\label{ssec:howToSelectA}
The recovery error occurs when $\sampleAfterTransformation_i$ has coarser precision than the original $\sample_i$ in the dataset. The new precision is shared by all $\sampleAfterTransformation_i$ when they have the same $E_U$ (See \autoref{fig:additionGuaranteedCommonBits}). The larger the $\A$ value, the more common bits the numbers share and the higher the recovery error. 
Let us illustrate this with an example. Consider $\sample_1 = 1.5$ and that $\A=10^7$. Then, $\sampleAfterTransformation_1=\sample_1 + \A = 10,000,002$  by adopting the \textit{round to nearest} rounding method from IEEE 754 (three other methods are supported \cite{FPHandbook}), since the precision of $\sampleAfterTransformation_1$ due to the $\A$ value being used is $P(E_U = 23) = 1.0$. After decompression, the recovered sample will be $\tilde{\sample}_1 = \sampleAfterTransformation_1 - \A = 2.0$ instead of $1.5$, generating an absolute error of $\errorAbs_1 = 2.0 - 1.5 = 0.5$.

\begin{figure}[t]
	\centering
	\includegraphics[width =0.7 \columnwidth]{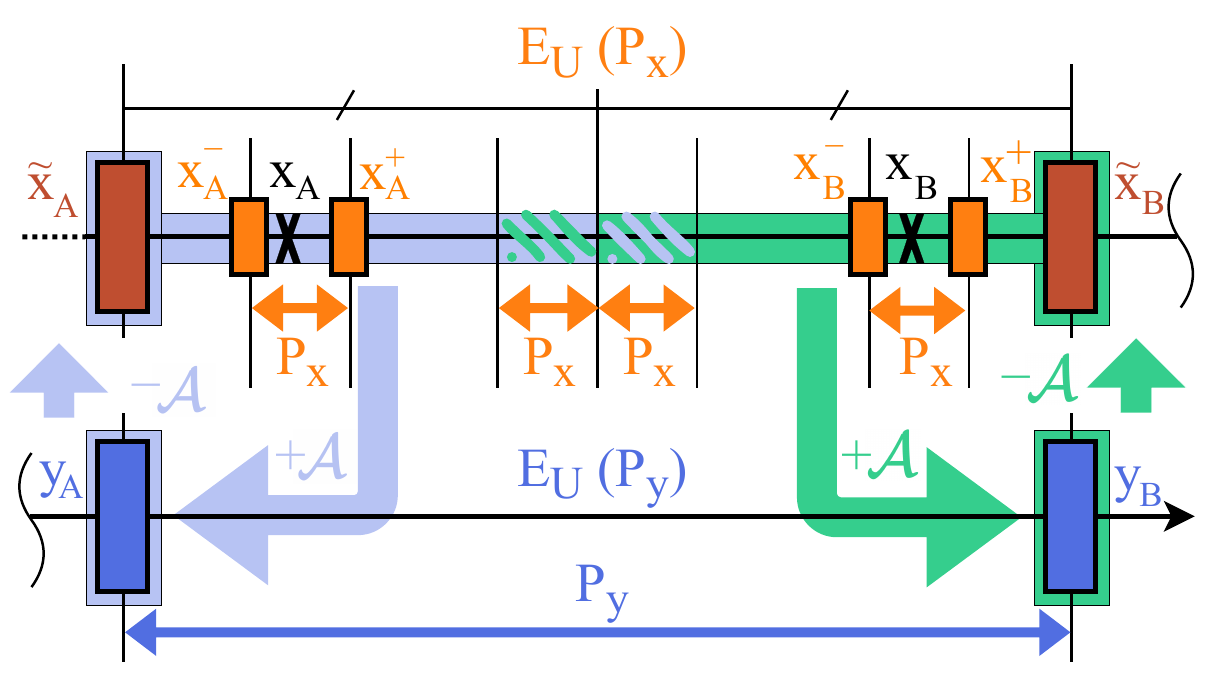}
	\caption{All $\sample_A \in \dataset$ in the violet region are transformed to $\sampleAfterTransformation_A$ and recovered with $\tilde{\sample}_A$. All $\sample_B \in \dataset$ in the green region go to $\sampleAfterTransformation_B$ and $\tilde{\sample}_B$. The result for samples from the region with both colors depends on $\A$ and $P_y$.}
	\label{fig:howApproximationWorks}
	\vspace{-0.5cm}
\end{figure}
\autoref{fig:howApproximationWorks} summarizes both the transform and its inverse starting from two real numbers $\sample_A, \sample_B\in  \mathbb{R}$, to their recovered form $\tilde{\sample}_A$ and $\tilde{\sample}_B$. Considering $\sample_A$, we use the rules from IEEE-754 to represent it with a 32 bits FP number and therefore store it as one of its closest neighbour $\sample_A^+, \sample_A^- $ in the set of representable FP numbers. Their precision is $P_x$. With the transform, we reach $\sampleAfterTransformation_A = \sample_A + \A $, having precision $P_y$. For ensuring more common mantissa bits, we choose $\A$ so that $P_y>P_x$. By applying the inverse transform for recovering the sample, we obtain $\tilde{\sample}_A= \sampleAfterTransformation_A - \A $. Both $\tilde{\sample}_A$ and $\tilde{\sample}_B$ belong to the subset of numbers in the $P_x$ region having no component smaller than $P_y$, since they have at least $\log_2(\frac{P_y}{P_x})$ zeros in the least significant portion of the mantissa. Supposing that all samples after being transformed have the same $E_U(P_y)$, $\tilde{\sample}_A$ and $\tilde{\sample}_B$ will be approximating every $\sample_i$ such that $\tilde{\sample}_A\leq \sample_i \leq \tilde{\sample}_B$. This is because given $\sample_i \in \dataset$, during the addition transform $\sampleAfterTransformation_{i} = \sample_i + \A $, we are losing any info smaller than $P_y$, and with $\tilde{\sample}_{i}= \sampleAfterTransformation_{i}  - \A $ we are filling the mantissa bits we lost with zeros. Therefore,
\begin{align}
\vspace{-0.7cm}
	\begin{split}
		\sample_i + \A &= \sampleAfterTransformation_A, \; \tilde{\sample}_A = \sampleAfterTransformation_A - \A \; \forall \sample_i \in [\tilde{\sample}_{A}; \tilde{\sample}_A + P_y/2 -P_x]\\
		\sample_j + \A &= \sampleAfterTransformation_B, \; \tilde{\sample}_B = \sampleAfterTransformation_B - \A \; \forall \sample_j \in [ \tilde{\sample}_{A} + P_y/2 + P_x; \tilde{\sample}_{B} ]
	\end{split}
\vspace{-0.7cm}
\end{align}
meaning that different numbers belonging to the same precision region will result in different recovery errors. Due to the standard, $\sampleAfterTransformation_{i}$ and $\tilde{\sample}_i$ for $\sample_i\in ] \tilde{\sample}_{A} + P_y/2 - P_x; \tilde{\sample}_{A} + P_y/2 + P_x [$ depend on the rounding method and the selected $\A$. The use of the addition transform ensures a bound in the recovery error, which is a function of $P_{\sampleAfterTransformation}$
\begin{equation}
	\errorAbs \leq 2^{E^{\sampleAfterTransformation}_U - 23 - 1} = \frac{P_{\sampleAfterTransformation} }{2},\\
	\label{eq:additionErrorBound}
\end{equation}
where $E^{\sampleAfterTransformation}_U$ is the unbiased exponent shared by all samples after the transform.

\subsubsection{Selecting the addition parameter}
We select $\A$ such that:
\begin{enumerate}
	\item[a.] it is as large as possible while keeping $\Delta$ and $\delta$ within the user requirements and complying with b) and c);	\label{Asela}
	\item[b.] the transformed dataset is aligned with the largest powers of 2 in the region;	\label{Aselb}
	\item[c.] it has the same precision of the numbers resulting from the addition with $\A$.	\label{Aselc}
\end{enumerate}
\begin{figure}[!t]
	\centering
	\vspace{0.1cm}
	\includegraphics[width = 0.7\columnwidth]{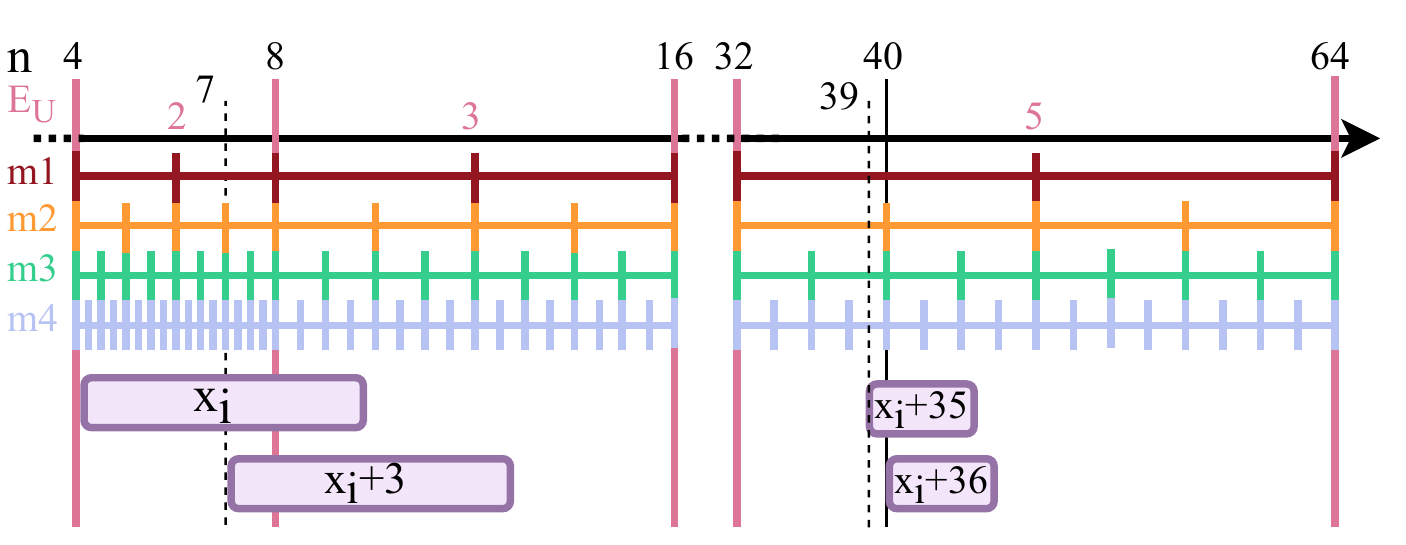}
	\caption{With \textcolor{violet}{violet}, the dataset range of values. $\A$ should be selected to make all samples share their exponent. Moreover, within the same precision region, some $\A$ result in more common bits than others.
	}
	\vspace{-0.3cm}
	\label{fig:alignWithPowersOf2}
\end{figure}
We can fulfil \hyperref[Asela]{principle [a.]} either by using the error bound in \autoref{eq:errorUpperBoundAbs} and \autoref{eq:errorUpperBoundRel} or by checking the real error for each $\sample_i \in \dataset$. Since we want all $\sampleAfterTransformation_i$ to share the exponent bits as well, we need to select $\A$ so that they lie on a region with equal precision. Considering the example in \autoref{fig:alignWithPowersOf2}, by choosing $\A = 3$, $E_{U}^{\sampleAfterTransformation_i} \in [2, 3]$, whereas with $\A = 35$, $E_{U}^{\sampleAfterTransformation_i} = 5 \; \forall \sampleAfterTransformation_i$.

\hyperref[Aselb]{Principle [b.]} comes from comparing the choice $\A = 35$ and $\A = 36$ in \autoref{fig:alignWithPowersOf2}. Although both shift the dataset to the same exponent region, $\A = 35$ will result in $m_2 \in\{0,1\}$, whereas with $\A = 36$, $m_2 = 1$, making $m_2 $ shared. The alignment with the powers of $2$ within the selected region is key, maximizing common mantissa bits. The larger the power of 2 aligned with the shifted dataset, the more mantissa bits are shared.

\hyperref[Aselc]{Principle [c.]} is needed to avoid a counterintuitive phenomenon while selecting $\A$.
Although conventional wisdom would suggest that the smaller we choose $\A$, the smaller and more precise the transformed numbers in the dataset are, ultimately leading to smaller recovery errors, this is not always the case: it is only true when $\sample_i + \A \in 2^{E_U^\A}$, where $E_U^\A$ is $\A$'s unbiased exponent. By selecting an $\A$ compliant with this last requirement, it is guaranteed that smaller addition parameters always lead to decreasing recovery errors. The implementation used in this paper for choosing $\A$ is
\begin{equation}
	\text{Select }\A : \max(\sample_i \in \dataset) + \A = 2^{E_U^\A + 1} - 2^{E_U^\A - 23}.
	\label{eq:AToFixLastNumberInDS}
\end{equation}

\subsection{Multiplication transform}
\label{sec:multiplicationMethod}

This method can be summarized as
\begin{equation}
	\sampleAfterTransformation_i = \hat{\sample}_i \cdot \M , \quad  \hat{\sample}_i = f(\sample_i, \M) \quad \forall {\sample}_i: \sample_i \in\dataset \\
	\label{eq:formulaMultiplicationMethod}
\end{equation}
where we call $\M$ the \emph{multiplication parameter}, and $f$ is a function that \textit{substitutes} each data sample $\sample_i$ in the dataset with $\hat{\sample}_i$, a version arbitrarily close to the original, but with the nice property of resulting in many zeros once multiplied by $\M$. So, in contrast to the addition transform, we apply the multiplication transform on $\hat{\sample}_i $ rather than on the original $\sample_i $. We aim to maximize the number of least significant bits in the mantissa equal to zeros. Let us explain this with an example.

\subsubsection{Numerical example}
\label{ssec:multiplicationMethodNumericalExample}
We consider $\sample_1 = 363.754$ and $\sample_2= 366$, with $\sample_2$ having $16$ consecutive zeros at the end of the mantissa and $\sample_1$ having none. In order to increase the number of common ending zeros, we propose substituting $\sample_1$ and $\sample_2$ with numbers that show the desired zeros after being multiplied by a carefully selected multiplication parameter $\M$. We consider $\hat{\sample}_1 = 363.7894592285156$ and $\hat{\sample}_2 = 366.03509521484375$ with $\M = 57$. Then, $\sampleAfterTransformation_1 = \hat{\sample}_1 \cdot \M  = 20736.0$ and $\sampleAfterTransformation_2 =\hat{\sample}_2 \cdot \M  = 20864.0$, which share the last $16$ zeros. By storing $\sampleAfterTransformation_1$, $\sampleAfterTransformation_2$ and $\M$, we can recover the original numbers via the divisions $\tilde{\sample}_1 =  \sampleAfterTransformation_1 / \M = \hat{\sample}_1 $ and $\tilde{\sample}_2 =   \sampleAfterTransformation_2 / \M = \hat{\sample}_2$ with maximum deviation $\errorPerc < 0.01\%$. Other existing methods, like \cite{infoPreprocessing}, directly approximate $\sample_1$ by removing all unwanted least significant mantissa bits to reach 16 ending zeros, where the recovered sample is $\tilde{\sample}_1 = 362.0$, with deviation $\errorPerc = 0.48\%$.


\subsubsection{Selecting the multiplication parameter}
\label{ssec:multiplicationMethodSingleNumber}
We use an approach that maps known values ($\M$) to known mantissa patterns that result in all zero sequences after their multiplication. A possible way to think about this is the equivalent in base 10. We know that all numbers ending with  $.5$, once multiplied by $\M = 2$, finish with $.0$, or that the ending sequence $1.25 \times 8$ always results in $0.00$. The same concept applies to floating point multiplication, where patterns expressed in base 2 can be exploited as substitutions for $\sample_i$'s mantissas. After $\sampleAfterTransformation_i = \hat{\sample}_i \cdot \M$, all bits in $\sampleAfterTransformation_i$'s mantissa, from the bit representing the starting power of 2 the pattern onward, will be zeros, making the length of the zeros-sequence an arbitrary choice.

\begin{table*}[!htbp]
	\caption{$\M$ unique patterns in base 2 (with 1 as MSB), expressed in hex}
	\label{tab:patterns}
	\centering
	\resizebox{\columnwidth}{!}{
	\begin{tabular}{Cc|Cc|Cc|Cc|Cc|Cc|Cc|Cc}
		\toprule
		$\M$ & Pattern & $\M$& Pattern & $\M$ & Pattern & $\M$ & Pattern & $\M$ & Pattern & $\M$ & Pattern & $\M$ & Pattern & $\M$ & Pattern \\ \midrule
		3 & 0x2 & 11 & 0x2e8 & 19 & 0x35e50 & 27 & 0x25ed0 & 35 & 0xea0 & 43 & 0x2fa0 & 51 & 0xa0 & 59 & 0x22b63cbeea4e1a0 \\
		5 & 0xc & 13 & 0x9d8 & 21 & 0x30 & 29 & 0x8d3dcb0 & 37 & 0xdd67c8a60 & 45 & 0xb60 & 53 & 0x9a90e7d95bc60 & 61 & 0x864b8a7de6d1d60 \\
		7 & 0x4 & 15 & 0x8 & 23 & 0x590 & 31 & 0x10 & 39 & 0xd20 & 47 & 0x572620 & 55 & 0x94f20 &  &  \\
		9 & 0x38 & 17 & 0xf0 & 25 & 0xa3d70 & 33 & 0x3e0 & 41 & 0xc7ce0 & 49 & 0x14e5e0 & 57 & 0x23ee0 &  & \\
		\bottomrule
	\end{tabular}
	}
	\vspace{-0.3cm}

\end{table*}
In \autoref{tab:patterns}, we list the unique patterns for each odd number $\M \in \{3,\dots,61\}$. The larger $\M$ is, the bigger the product's result and, therefore, the worse its precision, which generally leads to worse recovery errors. The table could be expanded by including for each $\M$ all patterns belonging to its factors, which also produce zeros. We do not consider even $\M$ since powers of 2 in multiplications only affect the exponent, leaving the mantissa intact.
\begin{table*}[ht!]
	\centering
	\caption{Examples of base patterns substitutions and multiplications}
	\resizebox{\columnwidth}{!}{
	\begin{tabular}{lllllllll}
		\toprule
		\textbf{$\sample_i = 19.6$} & \multicolumn{2}{l}{\textbf{$\M$ = }13}       & \multicolumn{1}{l}{\textbf{Pattern = }\boxForTable{000100111011} } &\multicolumn{3}{l}{\textbf{$\sample_i$ Mantissa} = 00111001100110011001101} &  &\\
		\toprule
		\textbf{$\hat{\sample}_i$} &
		\textbf{$E_U^{\hat{\sample}_i}$} &
		\multicolumn{1}{r}{\textit{Ext} }&
		\textbf{$\hat{\sample}_i$ Mantissa} &
		\textbf{$\sampleAfterTransformation_i$} &
		\textbf{$E_U^{\sampleAfterTransformation_i}$} &
		\textbf{$\sampleAfterTransformation_i$ Mantissa} &
		\textbf{Ending zeros} &
		\textbf{$\Delta$} \\
		\midrule

		19.61538506          & 4                 &                  & \noBoxForTable{0011}\boxForTable{\firstPattern{1}0011\underline{1}01100\underline{0}}\boxForTable{10011\underline{1}1} & 255.0 & 7 & 1111111\underline{\firstPattern{0}000000000000000} & 16 & 0.02 \\
		19.38461494          & 4                 &                  & \noBoxForTable{00}\boxForTable{\firstPattern{1}1\underline{01100}0\underline{1001}}\boxForTable{\underline{1}1\underline{01100}01} & 252.0 & 7 & 11111\underline{\firstPattern{0}00000000000000000 }& 18 & 0.22  \\
		18.46153831          & 4                 &                  & \boxForTable{\firstPattern{0}01\underline{0011}1\underline{0110}}\boxForTable{\underline{0}0\underline{10011}1\underline{01}1}& 240.0 & 7 & 111\underline{\firstPattern{0}0000000000000000000} & 20 & 1.14   \\
		29.53846169          & 4                 &          \multicolumn{1}{r}{{\boxForTable{\firstPattern{0}01}}} & \boxForTable{\underline{110}1100\underline{0}1}\boxForTable{0011\underline{1}0\underline{1}100\underline{0}1}\boxForTable{0\underline{1}} & 384.0 & 8 & 1\underline{\firstPattern{0}000000000000000000000} & 22 & 9.94   \\
		
		19.69230843          & 4                 &         \multicolumn{1}{r}{\boxForTable{\firstPattern{0}001}}& \boxForTable{001110\underline{1}1}\boxForTable{\underline{0}001\underline{0}0\underline{11}10\underline{1}1}\boxForTable{\underline{0}0\underline{1}} & 256.0 & 8 & \underline{\firstPattern{0}0000000000000000000000} & 23 & 0.09 \\
		\hline
		\multicolumn{9}{c}{\begin{minipage}{\textwidth}
				\vspace{1mm}
				$\sample_i \rightarrow \hat{\sample}_i \rightarrow \sampleAfterTransformation_i = \hat{\sample_i} \cdot \M $. Each pattern instance is boxed. Underlined bits differ from the originals, colored bit in the same row represent equal powers of 2. \textit{Ext} = mantissa extension: when looking for patches, we can append to the mantissa any number of 0s, followed by a 1.
			\end{minipage} 
		}
	\end{tabular}
	}
	\label{tab:floatingpointMultiplicationMethodExampleWithNumbers}
	\vspace{-0.3cm}
\end{table*}
Examples of this process are in \autoref{tab:floatingpointMultiplicationMethodExampleWithNumbers}, where $\sample_i = 19.6$ gets substituted in multiple ways using the pattern associated with $\M = 13$. Looking at the first row, suppose that we want to obtain $16$ ending zeros from a substitution of $19.6$ yet to be found after multiplying by $13$ and that our error bound imposes $E_U^{\sampleAfterTransformation_i} = 7$. Therefore, the zeros will start from $m_8$ (in red), which represents the power $2^{7-8} = 2^{-1}$. The bit in $\sample_i$ that corresponds to the same power is $m_5$ (in red), and we should start patching from that position onward. We use the bit-shifted version of the pattern minimizing necessary changes to the mantissa (underlined) to limit the error $\errorAbs$. Here, the pattern needs to be repeated twice (see boxes around the pattern) in order to fill the portion of the mantissa to be patched. After $m_{23}$, we crop it by rounding to the nearest, producing $\sampleAfterTransformation_i$.
From these examples we can conclude that:
\begin{itemize}
	\item Patterns are bit-shifting invariant.
	\item Starting patching from the left results in a larger error and more common zeros.
	\item Typically, the more bits you modify from the original mantissa, the bigger the error will be. 
	\item More zeros at the end of the mantissa do not necessarily mean a larger recovery error.
\end{itemize}

\subsubsection{Scaling up to datasets}

So far, we have applied the multiplication transform only to sets of up to 2 numbers, finding many possible substitutions with varying performances. However, we need to process larger collections of numbers, and looking at \autoref{fig:compressionDiagram} we want to find a single $\M$ that suits all $\sample_i$ and maximizes the number of common ending zeros. In the current implementation, we approach this problem with a brute force search, where first we analyze all $\sample_i \in \dataset$  $\forall \M\in [3, 61]$ looking for substitutions that fulfil the error bound. Among these multiplication parameters, we select the one that maximizes the number of common zeros for the dataset. The transformed dataset is then built by substituting each $\sample_i$ with $\hat{\sample}_i$ using the substitution corresponding to the chosen $\M$, having at least the minimum number of ending zeros found in the search and the smallest possible error.
\subsubsection{Multiply and check}
Due to the \textit{round to nearest} approximation method, there are some numbers $\tilde{\sampleAfterTransformation_i}$ that can not have $m_{23 }= 0$ when reached via $\tilde{\sampleAfterTransformation_i} = \hat{\sample_i} \cdot \M$. These numbers are off from being all-ending-zeros by $P$: after the multiplication, we might need to adjust the result with $\sampleAfterTransformation_i = \tilde{\sampleAfterTransformation_i} \pm P$.
\section{Performance evaluation}

In order to evaluate the performance of our proposed preprocessing techniques, we compare CR using the setups summarized in \autoref{tab:performanceSetup}.
\begin{table*}[!htbp]
	\centering
	\caption{Performance evaluation setup}
	\vspace{-0.1cm}
	\resizebox{\columnwidth}{!}{
	\begin{tabular}{l|l|ll}
		\toprule
		\textbf{Compressors} &\textbf{Preprocessing} &\textbf{Dataset collection} &\textit{Example}\\
		\midrule
		\colorbox{greedyColor}{\textcolor{greedyColor}{---}} \textit{Greedy-GD} \cite{GD_Greedy} &$\bigcirc$ - Addition transf. with $\errorBoundAbs$ and $\errorBoundPerc$  &	aarhus citylab \cite{AarhusKommune_2017}&[35.87] \\
		\colorbox{bzipColor}{\textcolor{bzipColor}{---}}  \textit{bzip2} \cite{bzip2}&$\bigtriangledown$ - Multiplication transf. with $\errorBoundAbs$ and $\errorBoundPerc$ &	chicago  \cite{CityChicago_water}  &[20.5,0,-0.082,0.055,2,12.7]\\
		\colorbox{lz4Color}{\textcolor{lz4Color}{---}}	\textit{lz4} \cite{lz4}  & $\times$ - Info content transf. with $\errorBoundAbs$ and $\errorBoundPerc$ \cite{infoPreprocessing}&	cbb g2 \cite{cbb} &[14310,403388] \\
		\colorbox{zstdColor}{\textcolor{zstdColor}{---}} \textit{zstd} \cite{Zstandard} &\multirow{2}{*}{\begin{tabular}[c]{@{}l@{}}\textcolor{lossless}{\textbf{-----}} - Lossless, it removes decimals by\\\hspace{0.5cm}multiplying all samples for a power of 10\end{tabular}}& cbb dim2 \cite{cbb} &[12856, 705226] \\
		& & uci (\textit{3500 samples}) \cite{uci} & [3.5892]\\
		&&cmummac (\textit{20000 samples})\cite{cmummac}& [-0.497314,0.425049,-0.036255,-0.755371]\\
		\bottomrule
	\end{tabular}
	}	
\vspace{-0.2cm}
\label{tab:performanceSetup}
\end{table*}
\begin{figure*}[!htbp]

	\centering
	\vspace{-0.6cm}
	\setkeys{Gin}{width=0.7\textwidth}
	\subfloat{\includegraphics{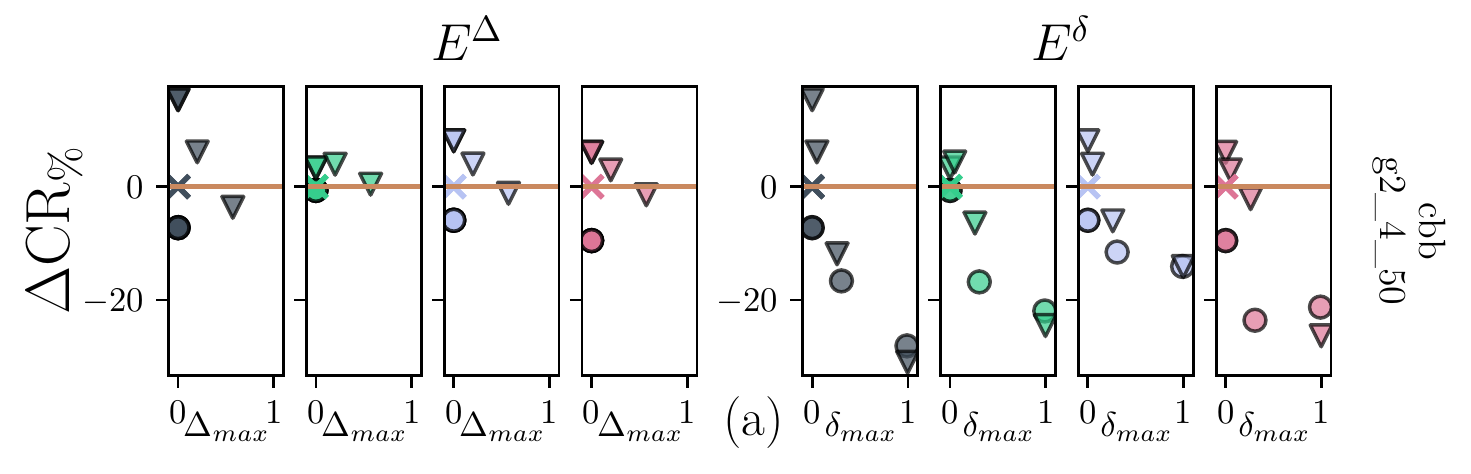}}\hfill
	\subfloat{\includegraphics{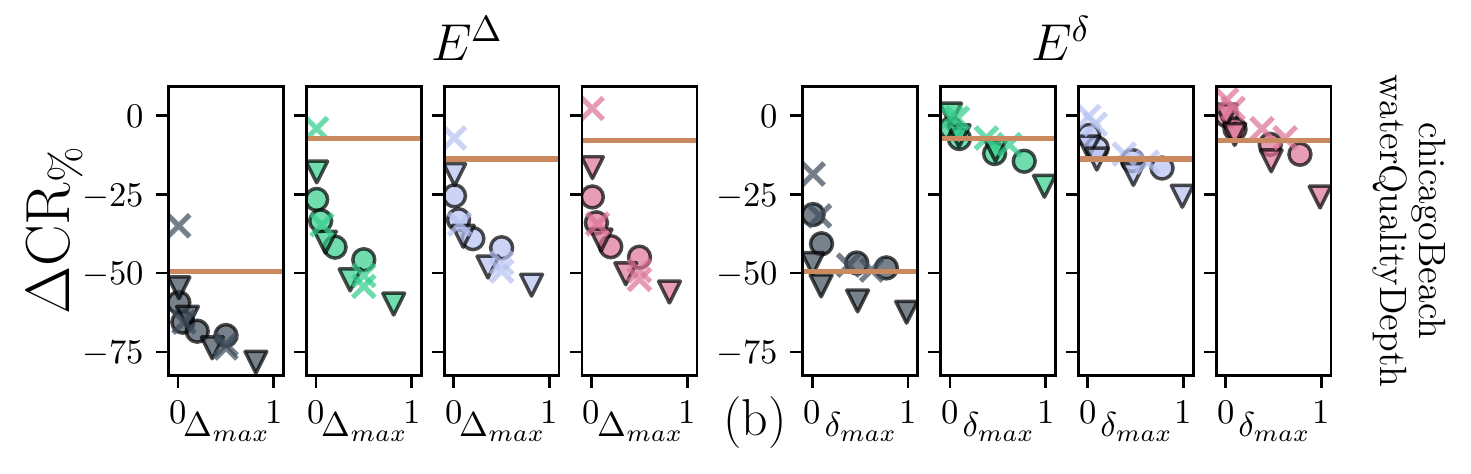}}\\
	\vspace{-0.5cm}
	\subfloat{\includegraphics{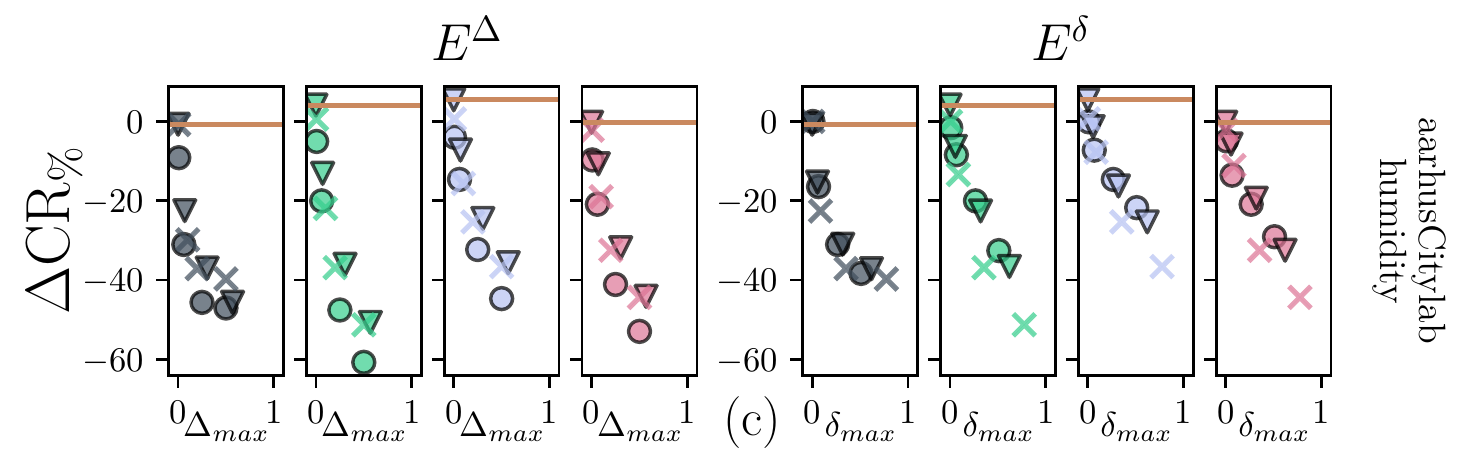}}\hfill
	\subfloat{\includegraphics{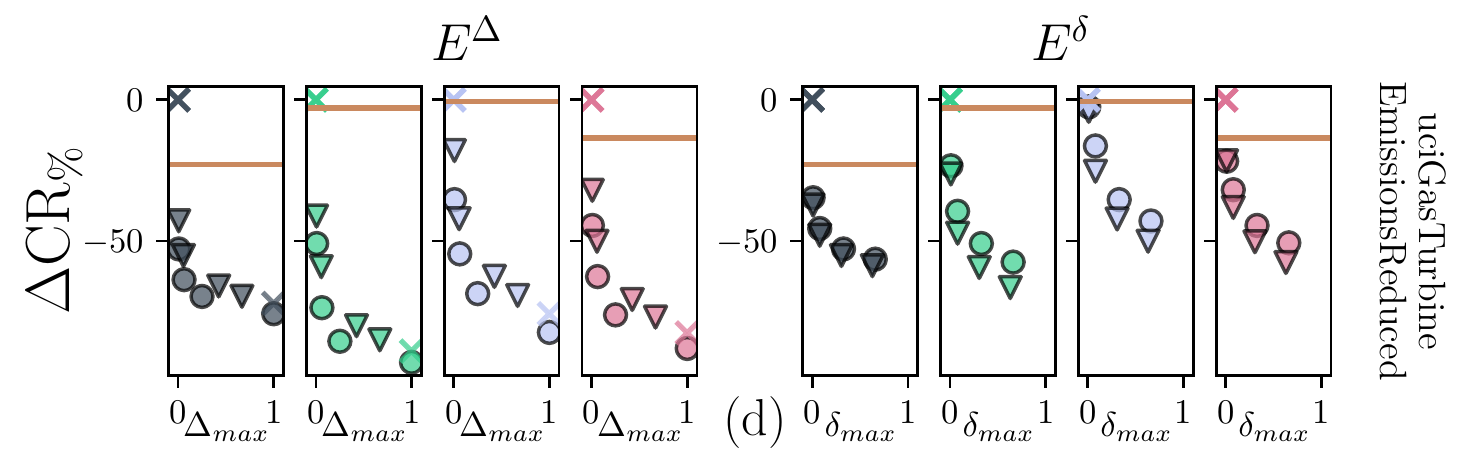}}\\
	\vspace{-0.5cm}			
	\subfloat{\includegraphics{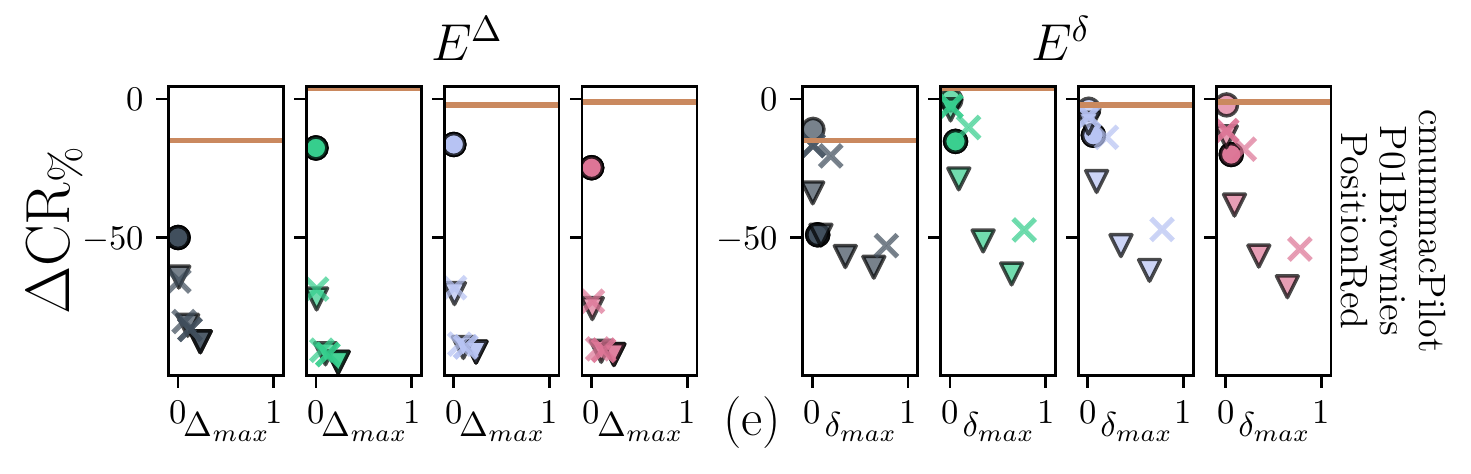}}\hfill
	\subfloat{\includegraphics{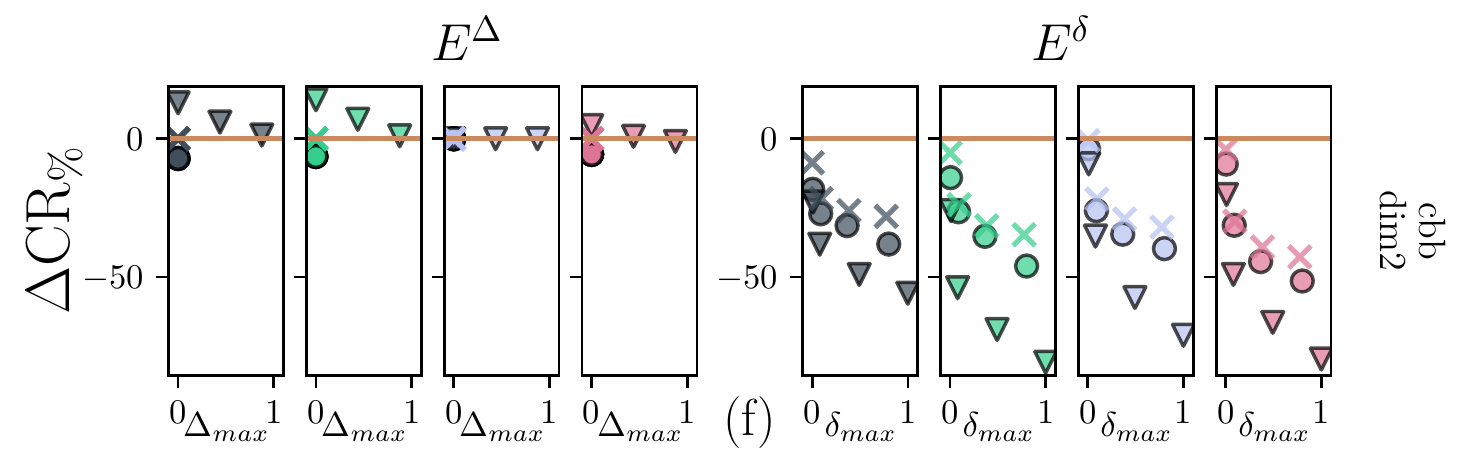}}\\
	
	\vspace{-0.3cm}
	\caption{Results to compare the novel preprocessing techniques against the lossy method in \cite{infoPreprocessing} ($\times$) and a lossless (orange line) solution, using an array of error bounds, compressors and datasets. All results are relative to the CR with no preprocessing, which acts as baseline $\Delta \text{CR}_\%  = 0$.}
	\vspace{-0.2cm}
	\label{fig:results}
\end{figure*}
We analyze the reduction in CR of our proposed addition and multiplication transforms, against a similar lossy preprocessing technique and a lossless method. The recent work \cite{infoPreprocessing} also tries to maximize the number of ending zeros in the mantissa to achieve better compression. It does so by computing the information content of each mantissa bit, rounding to zero the ending mantissa bits below a certain threshold. In order to adapt this algorithm to our error-bound approach, for all dimensions in each dataset, we found the minimum retained information content limit that fulfilled the error condition under analysis. The \textit{lossless preprocessing method }simply eliminates all decimals in the dataset by multiplying all samples with a power of 10.

We use the no-preprocessing results ($\text{CR}_\text{NP}$) as baseline for comparing performances, with the formula
\begin{equation}
\Delta \text{CR}_\% = ( \text{CR} - \text{CR}_{\text{NP}})/\text{CR}_{\text{NP}} \cdot 100.
\end{equation}
Therefore, $\Delta \text{CR}_\% < 0$ indicates improved compression. In \autoref{fig:results}, on the horizontal axis, we plot the actual maximum error on the recovered dataset imposing the error bounds $\errorBoundAbs \in \{0.01, 0.1,0.5, 1.0\}$ and $\errorBoundPerc \in \{0.01\%, 0.1\%,0.5\%, 1.0\%\}$. The chosen datasets represent different combinations of datatypes, sizes and number of dimensions: \textit{aarhus citylab} \cite{AarhusKommune_2017} and \textit{uci} \cite{uci} for floats with one dimension, \textit{cmummac} \cite{cmummac} for multidimensional floats, \textit{cbb gs} and \textit{cbb dim2 }\cite{cbb} for multidimensional integers and \textit{chicago} \cite{CityChicago_water} for a mix of integers and floats. The compressors we use are \textit{Greedy-GD}\cite{GD_Greedy}, a dictionary-based algorithm using deduplication with random access capabilities, as well as three other well-established compressors for comparison, namely \textit{bzip2} \cite{bzip2}, \textit{LZ4} \cite{lz4} and \textit{Zstandard} \cite{Zstandard}.

The results in \autoref{fig:results} show that our proposed preprocessing methods outperform the one in \cite{infoPreprocessing} for all datasets and compressors except in \autoref[c]{fig:results} using $\errorBoundPerc$, achieving up to 46\% better CR in \autoref[f]{fig:results}. In \autoref[a]{fig:results} and \autoref[d]{fig:results}, the information content transform cannot achieve errors below the desired bounds, leaving the datasets unprocessed with $\Delta$CR$_\% = 0$.

For all datasets and compressors, with $\Delta \leq 1$ or $\delta \leq 1\%$, both addition and multiplication transforms improve compression with respect to the non-preprocessed datasets, with reductions up to $80\%$. We see cases in \autoref[a]{fig:results}, \autoref[c]{fig:results} and \autoref[f]{fig:results} with $\errorBoundAbs$ where for small error bounds the non-preprocessed dataset performs better: however, by switching from $\errorBoundPerc$ to $\errorBoundAbs$ (and vice-versa) or relaxing a bit the error bound, we always achieve improvements. 

We also notice that when addition and multiplication transforms outperform the non-preprocessed datasets, they surpass the lossless performances too, represented by the orange line. This line lies on $y = 0$ in \autoref[a]{fig:results} and \autoref[f]{fig:results} since they are integer datasets: the lossless algorithm has no effect on them. As expected, a higher compression can be achieved at the expense of higher recovery errors.

Moreover, the choice between addition and multiplication transform depends on the dataset and the error-bound method. For example, in \autoref[c]{fig:results} with $\errorBoundAbs$, the addition outperforms multiplication, while in \autoref[f]{fig:results} with $\errorBoundPerc$ multiplication is better.

Regarding the error bounds, the choice between $\errorBoundAbs$ and $\errorBoundPerc$ also depends on the dataset. In \autoref[f]{fig:results}, choosing a relative error bound is clearly more effective than an absolute one, while in \autoref[b]{fig:results} the absolute bound produces better results. Generally, $\errorBoundAbs$ is more suited for datasets having values close to zero. We should be careful at selecting $\errorBoundAbs$ so that it does not compromise the information carried by the dataset. An example of this in \autoref[e]{fig:results}, where $\errorBoundAbs$ can produce reductions close to $95\%$: however, as we can see from the examples in \autoref{tab:performanceSetup} for dataset \textit{cmummac}, selecting $\errorBoundAbs = 1$ would make the recovered dataset unusable. Therefore, in this case we should opt for $\errorBoundPerc$.

\section{Conclusions}
In this paper, we proposed two novel lossy preprocessing techniques to improve the compression ratio of existing compressors under given error bounds by transforming the dataset before compressing it. These two methods use simple floating point arithmetic operations like addition and multiplication, as well as floating point data structure, to increase the number of common bits throughout the whole dataset. We presented the performances of these techniques by comparing their resulting CR against the one obtained without preprocessing, with lossless preprocessing, and using a similar lossy preprocessing technique \cite{infoPreprocessing}, considering four compressors and six datasets. We plan to assess extensions of these ideas into lossless methods, as well as to achieve improvements in terms of preprocessing time.
\label{sec:conclusions}
\bibliographystyle{IEEEtran}
\bibliography{IEEEabrv,library,IEEEtran_control}

\end{document}